\documentclass[%
reprint,
twocolumns,
aps,
pre,
amsmath,
amssymb,
showpacs,
notitlepage,
nobibnotes,
superscriptaddress]
{revtex4-2}
\usepackage{graphicx} 
\usepackage[colorlinks=true,linkcolor=blue,citecolor=blue,filecolor=cyan,urlcolor=blue,breaklinks=true]{hyperref}
\usepackage{mathtools}
\usepackage{amsmath}
\usepackage{subcaption}
\usepackage[font=footnotesize,labelfont=bf,justification=raggedright]{caption}

\begin{document}

\title{Understanding random-walk dynamical phase coexistence through waiting times}

\author{David C. Stuhrmann
}
\affiliation{Department of Physics, Stockholm University, Roslagstullsbacken 21, SE-106 91 Stockholm, Sweden}

\author{Francesco Coghi
}
\email{francesco.coghi@su.se}
\affiliation{Nordita, KTH Royal Institute of Technology and Stockholm University, Hannes Alfvéns väg 12, SE-106 91 Stockholm, Sweden}

\date{\today}

\begin{abstract}
We study the appearance of first-order dynamical phase transitions (DPTs) as `intermittent' co-existing phases in the fluctuations of random walks on graphs. We show that the diverging time scale leading to critical behaviour is the waiting time to jump from one phase to another. This time scale is crucial for observing the system's relaxation to stationarity and demonstrates ergodicity of the system at criticality. We illustrate these results through three analytical examples which provide insights into random walks exploring random graphs.
\end{abstract}




\maketitle


\textit{Introduction.} Random walks are arguably the most versatile model~\cite{Grimmet2001,Gregory2010} to describe various transport processes in natural and artificial environments described as complex networks, including spreading of infections, propagation of information, search algorithms, and community detection~\cite{Hughes1995,Noh2004,Yang2005,Barrat2008,Newman2010,Masuda2017}. Often, studies focus on dynamical observables whereby a single random walker hops on a graph and accumulates information related to some characteristics of the states visited over time. Among these dynamical observables we list first-passage times, currents, occupation times and entropy production rates. Although the typical behaviour of these dynamical observables obtained for long times is nowadays of easy interpretation, fluctuations and rare events, which are important over finite observation times, are less understood. In particular, among all possible rare events, dynamical phase transitions are strongly relevant.


Dynamical phase transitions (DPTs) are considered as changes in the random-walk fluctuation mechanisms. These have been observed in various models of single and many-particle interacting systems, involving the limit of certain parameters~\cite{WhitelamComment}. These are large system sizes~\cite{Bodineau2005,Bertini2005,Garrahan2007,Vaikuntanathan2014,DeBacco2016,Shpielberg2016,Baek2017,Baek2018,Whitelam2018,Coghi2019,Buca2019,Whitelam2021,Carugno2022,Gutierrez2021a}, large particle numbers~\cite{Nemoto2019,Agranov2023} and system volumes~\cite{Bunin2012,Proesmans2020}, or parameters associated to stochastic resetting~\cite{Harris2017,Zamparo2019a,Coghi2020,Zamparo2022,Mori2022,Zamparo2023}, or again small rate/weak noise limits~\cite{Baek2015,TsobgniNyawo2016,Proesmans2019,DiGaetano2023} and strong driving fields~\cite{Espigares2013} of driven particle models. Recently, DPTs were observed in association with anomalous scaling of large deviations too~\cite{Nickelsen2018,Smith2022,Stella2023}.

DPTs are considered to arise whenever a non-analytic (critical) behaviour appears in large deviation functions, e.g., scaled cumulant generating functions (SCGFs) or rate functions. This critical behaviour is surely a necessary condition, unlike time-reversal symmetry breakings~\cite{Bunin2012,Shpielberg2016,Baek2017,Baek2018}, manifesting the divergence of a relevant time-scale in the system. In the context of Markov processes, such a diverging time scale may arise as consequence of metastability~\cite{Larralde2005,Larralde2006,Macieszczak2016,Macieszczak2021} whereby the system slowly relaxes from a state that appears stationary at short times towards another that is genuinely stationary. 

Nonetheless, it is important to note that a non-differentiable point in a large deviation function is not enough to demonstrate a transition, such as phase coexistence, in the fluctuations of a dynamical observable. In systems that possess a well-defined free energy and adhere to the Landau theory of phase transitions, such as equilibrium systems, this issue does not arise. However, when examining the time-dependent dynamics of a system and considering a large deviation function, the latter might not exhibit all the characteristics of a free energy, except in several cases as demonstrated in~\cite{Baek2015,Baek2017,Baek2018}, despite sharing the same mathematical form. Therefore, in such a scenario, in order to link a kink in a SCGF to a physical transition one needs to better understand the phenomenology of the model being analysed~\cite{Whitelam2018,WhitelamComment,Whitelam2021}, which may or may not show equilibrium-like phases.


In this paper, we show that when a non-differentiable point appears in the SCGF of a time-additive observable of a random walk, the relevant diverging time scale---akin to a correlation length in equilibrium statistical mechanics---supporting the co-existence of two phases is the waiting time to jump from a phase to the other. Having two co-existing phases is intended at the level of single random-walk trajectories being `intermittent', i.e., the random walker keeps hopping from a phase to the other. In such a scenario, we will show that by opportunely re-scaling the large deviation functions with the jump waiting time, the kink in the SCGF disappears, restoring a fully analytic function and the large deviation principle for the observable under examination. This means that the system once observed with the right time scale is still ergodic and therefore the walker visits all regions of the state space. We illustrate our findings through three analytical examples. Two of these are borrowed from a previous publication~\cite{Carugno2022}, while one is entirely original. We believe these examples help to understand fluctuations of random walks exploring random graphs~\cite{DeBacco2016,Coghi2019,Gutierrez2021}, as described toward the end of the paper.

\textit{Model set up, large deviations, and the driven process.} We consider an $n$-time-step random walk (RW) $X = (X_1, X_2, \cdots, X_\ell, \cdots, X_n)$ on a finite connected and undirected graph $G=(V,E)$ characterised by a set of states $V$ of finite size $N$ and a set of edges $E$. The RW dynamics is determined by the $N \times N$ stochastic transition matrix $\Pi = \left\lbrace \pi_{ij} \right\rbrace$ giving the probability $\pi_{ij}$ of the RW to move from $X_\ell = i$ at time $\ell$ to $X_{\ell+1} = j$ at time $\ell+1$. If $i$ and $j$ are not connected $\pi_{ij} = 0$, otherwise $0 < \pi_{ij} \leq 1$.

At each time step $\ell$ the RW collects a certain `cost' (or `reward', `observable') related to the state of the system visited, namely $f(X_\ell)$. We focus on the dynamical observable that characterises the mean cost visited by a random walk trajectory, 
\begin{equation}
\label{eq:Obs}
C_n = \frac{1}{n} \sum_{\ell = 1}^n f(X_\ell) \ ,
\end{equation}
with $f$ a bounded function. Because the RW is ergodic---this is guaranteed by the properties of the graph $G$~\cite{Touchette2009,Dembo2010}---$C_n$ converges to the ergodic average
\begin{equation}
\label{eq:Ave}
\sum_{i \in V} \rho_i f(i) \coloneqq c^* \ ,
\end{equation}
where $\rho = \{ \rho_i \} $ is the stationary distribution of the RW. For a finite observation time, the observable $C_n$ is a random variable of the random walk process and its distribution $P_{N,n}(C_n=c) \coloneqq P_{N,n}(c)$ is known to take the large deviation form~\cite{DenHollander2000,Touchette2009,Dembo2010}
\begin{equation}
\label{eq:Rate}
P_{N,n}(c) = e^{-n I_N(c) + o(n)} \ ,
\end{equation}
with the time-leading behaviour given by the non-negative large deviation rate function $I_N(c)$ and $o(n)$ denoting corrections smaller than linear in $n$. 

The large deviation rate function $I_N(c)$ can be hard to derive if the distribution $P_{N,n}(c)$ is unknown. In such a case, the rate function can be calculated by means of the so-called G\"{a}rtner--Ellis theorem~\cite{DenHollander2000,Touchette2009,Dembo2010}. This states that given a differentiable scaled cumulant generating function (SCGF)
\begin{equation}
\label{eq:SCGF}
\Psi_N(s) = \lim_{n \rightarrow \infty} \frac{1}{n} \ln \mathbb{E} \left[ e^{n s C_n} \right] \ ,
\end{equation}
the rate function $I_N$ can be obtained via the following Legendre transform:
\begin{equation}
    \label{eq:LegendreFenchel}
    I_N(c) = s^* c - \Psi_N(s^*) \, ,
\end{equation}
where $s^*$ is the unique root of $\Psi_N'(s) = c$~\cite{Touchette2009}.
In particular, given that the RW $X$ is ergodic, the form of the SCGF simplifies to
\begin{equation}
\label{eq:SpectralSCGF}
\Psi_N(s) = \ln \zeta_s ,
\end{equation}
where $\zeta_s$ is the dominant eigenvalue of the so-called tilted matrix $\tilde{\Pi}_s = \left\lbrace \left( \tilde{\pi}_s \right)_{ij} \right\rbrace$, with components
\begin{equation}
\label{eq:TiltedMatrix}
\left( \tilde{\pi}_s \right)_{ij} = \pi_{ij} e^{s f(i)} \ .
\end{equation}

The large deviation picture is complete once we understand how fluctuations $C_n = c$ are created in time. To do so, we construct the driven process~\cite{Chetrite2013,Chetrite2015,Chetrite2015a} which, in this context, is a biased RW~\cite{Chetrite2015,Coghi2019} whose transition probability matrix is given by
\begin{equation}
\label{eq:DrivenProcess}
\left( \pi_s \right)_{ij} = \frac{\left( \tilde{\pi}_s \right)_{ij} r_s(j)}{r_s(i) e^{\Psi_N(s)}} \ ,
\end{equation}
where $r_s$ is the right eigenvector associated with $\zeta_s$. Under this process, the observable $C_n$ is asymptotically distributed according to the canonical form
\begin{equation}
    \label{eq:CanonicalDriven}
    P^{(s)}_{N,n}(C_n=c) = \frac{e^{n s c} P_{N,n}(c)}{\mathbb{E} \left[ e^{n s C_n} \right]} \ .
\end{equation}
The driven process is well defined for finite $N$, still Markovian, and ergodic and can be interpreted as the effective dynamics of the subset of paths of the original RW leading to a fluctuation $C_n = c = \Psi_N'(s)$~\cite{Coghi2019,Gutierrez2021}.

\textit{Dynamical phase transitions: phase coexistence.} Thanks to the Perron--Frobenious theorem, at finite $N$, the SCGF \eqref{eq:SCGF} and the rate function \eqref{eq:LegendreFenchel} are both analytic functions~\cite{Touchette2009,Dembo2010}. However, there is no general theorem that guarantees that for infinite $N$
\begin{equation}
    \label{eq:LimitN}
    \Psi(s) \coloneqq \lim_{N \rightarrow \infty} \Psi_{N}(s)  \, ,
\end{equation}
and its Legendre transform $I(c)$ are analytic. As mentioned in the introduction, there are many cases described in the literature where large deviation functions show singular points for such a limit.

Often, these singularities are interpreted as DPTs, viz.\ changes in the mechanisms that generate particular fluctuations of the observable $C_n$. In studying transitions, we follow a common practice used for equilibrium systems, as proposed by Ehrenfest~\cite{Ehrenfest1933}. Even for time-dependent models, we use the non-continuous derivative of the SCGF to determine the order of the transition. \footnote{Sometimes scientists define the order of the DPT by looking at the rate function rather than the SCGF. The SCGF $\Psi(s)$ is, arguably, closer in form to a Helmoltz (canonical) free energy although this, strictly speaking, would be obtained by dividing $-\Psi(-s)$ by the Laplace parameter $s$~\cite{Touchette2009}. We do not do that because $\Psi(s)$ is already, conveniently, convex. Furthermore, because of the Legedre transform \eqref{eq:LegendreFenchel} connecting the SCGF to the rate function, a DPT of order $n$ in the SCGF will be interpreted as of order $n+1$ in the rate function.}. For this reason, a non-differentiable point $s_c$ in the SCGF $\Psi(s)$ is often referred to be a first-order DPT and therefore to signal an abrupt change in the fluctuations of $C_n$ and, consequentially, the emergence of phase coexistence. However, it has recently been pointed out that such singular response of the SCGF is not strictly related with coexisting phases.

Indeed, other scenarios could arise, such as that of a slow system, or a pure ergodicity-breaking transition~\cite{Whitelam2018,Whitelam2021}. Applied to our context, in the former case, the random walker is simply extremely slow in moving from a phase to the other and the overall picture is that one of a RW slowly leaving a metastable state and being absorbed by a stable one. In the latter, we assist to the breaking of the large deviation principle for the observable under examination and therefore trajectories either visit one phase or the other (no mixing). Albeit their different physical interpretations, all these scenarios are commonly described by the blow-up of a characteristic time scale, namely $\tau(N)$, for $N \rightarrow \infty$.

Fully characterising the time scale $\tau(N)$ allows one to properly re-scale the Laplace parameter $s$ around $s_c$, the SCGF $\Psi(s)$, and the rate function $I(c)$ as
\begin{align}
    \label{eq:Ress}
    \bar{s}(s) &= \tau(N) (s + s_c) \\
    \label{eq:ResPsi}
    \bar{\Psi}_N(\bar{s}) &= \tau(N) \left( \Psi_N(s(\bar{s})) - \Psi(s_c) \right) \\
    \label{eq:ResI}
    \bar{I}_N(c) &= \tau(N) I(c) \, ,
\end{align}
such that $\bar{s}$ and $\bar{\Psi}_N(\bar{s})$ are now centered around $0$ and we can re-write \eqref{eq:Rate} at leading order in $n$ and for an observation time $n \gg \tau(N)$ \footnote{This, in turn, explains the order of the limits taken, which often do not commute.} as $P_{N,n}(c) \approx e^{-\bar{n} \bar{I}_N(c)}$ with a new `speed' given by $\bar{n} \coloneqq n/\tau(N)$. The system restores a large deviation principle and does not experience ergodicity breaking if the functions \eqref{eq:ResPsi} and \eqref{eq:ResI} are smooth for $N \rightarrow \infty$.

Incidentally, from \eqref{eq:CanonicalDriven} we can obtain the rate function and, by Legendre transform, the SCGF of the driven process which read
\begin{align}
    \label{eq:DrivRate}
    I^{(s^*)}_N(c) &= I_N(c) - s^* c + \Psi_N(s^*) \\
    \label{eq:DrivPsi}
    \Psi^{(s^*)}_N(s) &= \Psi_N(s+s^*) - \Psi_N(s^*) \, ,
\end{align}
having replaced $s$ with $s^*$ in \eqref{eq:CanonicalDriven}. Therefore, $\bar{\Psi}_N(\bar{s})$ in \eqref{eq:ResPsi} and $\bar{I}_N(c)$ in \eqref{eq:ResI} represent the $\tau(N)$-rescaled versions, for $s^* = s_c$, of \eqref{eq:DrivPsi} and \eqref{eq:DrivRate} respectively. In other words, the function $\bar{I}(c)$ is the limiting (for $N \rightarrow \infty$) rate function associated with the observable $C_n$ of the driven process at $s_c$, opportunely rescaled by the diverging time scale $\tau(N)$. The latter provides an indication of the required simulation time for the driven process to relax and yield reliable statistics of $C_n$.

Finally, we argue that in the case of a two-phase coexistence, signalled by a non-differentiable point in the SCGF $\Psi(s)$, the relevant time scale $\tau(N)$ is played by the waiting time to jump from a phase of the system to the other. In the following, we discuss three illustrative examples of RWs exploring graphs. For these, phase coexistence in the fluctuations of $C_n$ is visualised by simulating the driven process in the vicinity of the critical parameter $s_c$ for large values of the parameter $N$ and checking that RW trajectories are intermittent between the two phases (see, for instance, Fig.\ \ref{fig2:2int}).

\textit{Example 1: Two-state Random Walk.} We start by considering a two-state RW. We name the two states `chain' and `bulk' and define the transition matrix to be 
\begin{equation}
    \label{eq:2StateTrans}
    \Pi = 
    \begin{pmatrix}
    \frac{1}{2} & \frac{1}{2} \\
    \frac{1}{N} & 1-\frac{1}{N}
    \end{pmatrix} \, ,
\end{equation}
such that, by increasing $N$, the RW spends on average more time in the `bulk'. [Notice that the parameter $N$ here is not the size of the graph (as considered when setting up the model). However, it will play the same role.] Finally, the observable considered is~\eqref{eq:Obs} with $f(\text{chain}) = 1/N$ and $f(\text{bulk}) = 1$. 

\begin{figure}[h!]
    \centering
\includegraphics[width=\linewidth]{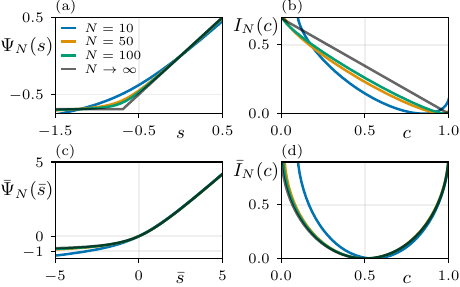}
    \caption{Two-state RW: SCGFs $\Psi_N(s)$ in (a) and rate functions $I_N(c)$ in (b) at increasing values of $N$ (coloured curves) along with their limiting functions $\Psi(s)$ and $I(c)$ (dark grey curves). Re-scaled versions $\bar{\Psi}_N(\bar{s})$ and $\bar{I}_N(c)$ (coloured curves), along with their limits $\bar{\Psi}(\bar{s})$ and $\bar{I}(c)$ (dark grey curves) in (c) and (d).}
    \label{fig1:2scgf}
\end{figure}

\begin{figure}[h!]
    \centering
\includegraphics[width=\linewidth]{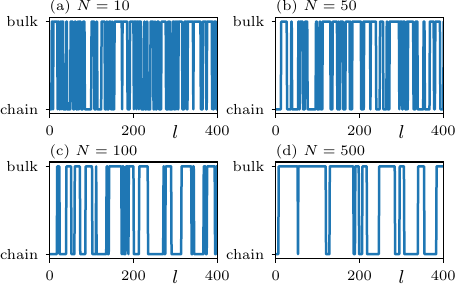}
    \caption{From (a) to (d), intermittent trajectories visiting `bulk' and `chain' at increasing values of $N$ for the two-state RW model. The larger the $N$ the longer the waiting time to visit the other phase.}
    \label{fig2:2int}
\end{figure}

Typically, according to \eqref{eq:Ave}, the RW collects a cost $c^* = (2 + N^2)/(2 N + N^2)$ which approaches $1$ for large $N$. 
SCGF~\eqref{eq:SCGF} and rate function~\eqref{eq:Rate} characterising the long-time fluctuations of this observable have already been studied in~\cite{Carugno2022}. In the limit $N \rightarrow \infty$ the SCGF $\Psi(s)$ develops a kink at $s = s_c \coloneqq - \ln 2$ which leads to a linear section in $I(c)$ (see black curves in the panels (a) and (b) of Fig.\ \ref{fig1:2scgf}). For $s < s_c$, fluctuations arise from a longer time spent in `chain', while for $s > s_c$, fluctuations arise from a longer time spent in `bulk'.

At the critical value $s_c$, the walker will spend half its time in `chain' and the other half in `bulk'. In particular, in the long-time limit an intermittent behaviour will arise whereby the walker, after spending a certain amount of time in `chain', will hop onto `bulk' and vice-versa. For such a simple model, there is no other way the RW can create the fluctuation associated with $s_c$. We capture this intermittent behaviour in Fig.\ \ref{fig2:2int} by simulating the driven process~\eqref{eq:DrivenProcess} with $s = s_c$ for finite but increasingly larger values of $N$.

Evidently, the larger the $N$ the longer the time the RW waits before hopping to the other phase. This is the relevant---arguably the only for this model---time scale $\tau(N)$ that diverges for $N \rightarrow \infty$ and that leads to the appearance of a kink in the SCGF at $s_c$. The derivation of this effective time scale was accomplished in a previous work~\cite{Carugno2022}---without much physical insights---by considering a general power-law form of $\tau(N)$, analytically expanding \eqref{eq:ResPsi} in $N$ and selecting the correct exponent for the power law such that the leading order of the expansion $\bar{\Psi}(\bar{s}) \coloneqq \lim_{N \rightarrow \infty} \bar{\Psi}_N(\bar{s})$ is not trivial. 
For a detailed explanation of the derivation, we refer the reader to~\cite{Carugno2022}. In the bottom panels (c) and (d) of Fig.\ \ref{fig1:2scgf} we display $\bar{\Psi}_N(\bar{s})$, its Legendre transform $\bar{I}_N(c)$ and their respective limits $\bar{\Psi}(\bar{s}) = \bar{s} + \sqrt{4 + \bar{s}^2} - 2$ and $\bar{I}(c)$. Because of the smoothness of these last functions, in the time scale defined by $\tau(N)$ a large deviation principle is restored and fluctuations can be studied at $s_c$.

\begin{figure*}[ht] %
    \includegraphics[width=\linewidth]{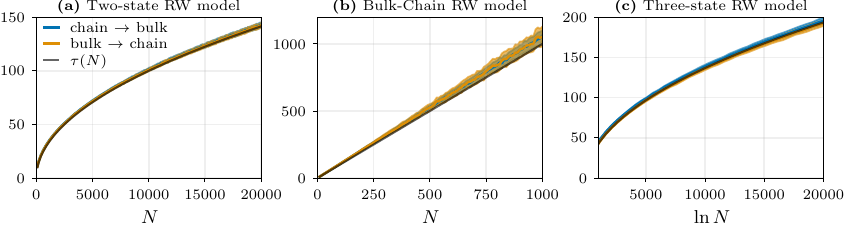}
    \caption{Time re-scalings $\tau(N)$ for the models investigated (black solid lines) compared with mean waiting times---in (c), these are multiplied by the $N$-independent prefactor $1.37$---to hop from one phase to the other for simulations of driven processes at $s=s_c$. For each $N$ we run $100$ simulations of $n=10^6$ time steps (average marked as a solid coloured line within one standard deviation).}
    \label{fig3:mwt}
\end{figure*}

In Fig.\ \ref{fig3:mwt}(a) we plot $\tau(N) = \sqrt{N}$ and compare it with the mean waiting time to hop from `chain' to `bulk' and vice-versa at increasing values of $N$. The average value of this quantity is obtained by simulating a driven process at $s=s_c$, counting the transitions unidirectionally between the states, and dividing the simulation time by the number of transitions. Furthermore, we remark that $\tau(N)$ has the same scaling form of the relaxation time of the driven process for $s=s_c$ calculated as the negative inverse of the spectral-gap logarithm of \eqref{eq:DrivenProcess} (not displayed in Fig.\ \ref{fig3:mwt}(a) as fully overlapping $\tau(N)$---modulo $N$-independent prefactors). Evidently, $\tau(N)$ well matches numerical simulations providing evidence that the critical time scale in a phase-coexistence scenario is determined by the waiting time between jumps.

\textit{Example 2: Bulk-Chain Random Walk.} We now focus on a slightly more complicated model. Differently from the previous case, the fluctuation at the critical point can arise through two different mechanisms and we will show that the dominant one is an intermittent behaviour supporting phase coexistence. 

We analyse an unbiased RW with transition matrix 
\begin{equation}
    \label{eq:BulkDanglingTrans}
    \pi_{ij} = \frac{a_{ij}}{k_i} \, ,
\end{equation}
with $A = \left\lbrace a_{ij} \right\rbrace$ representing the adjacency matrix of a graph of $N$ nodes, i.e., $a_{ij}=1$ if states $i$ and $j$ are connected and $a_{ij}=0$ otherwise, and $k_i = \sum_{j=1}^N a_{ij}$ representing the connectivity of state $i$. The graph is composed by a fully connected bulk of $N-2$ nodes and a single chain of $2$ nodes with connectivity $2$ and $1$ (we name this structure `dangling chain'). Because of symmetry, the graph has only $4$ qualitatively-different nodes. These are: the node of degree $1$ at the edge of the chain, the node of degree $2$ in the middle of the chain, the node of degree $N-2$ (gateway from now on) connecting bulk and chain, and a representative node of the bulk of degree $N-3$. On such a structure, the unbiased RW collects a cost of the form \eqref{eq:Obs} with $f(X_\ell) = k_{X_\ell}/N$.

\begin{figure}
    \centering
    \includegraphics[width=\linewidth]{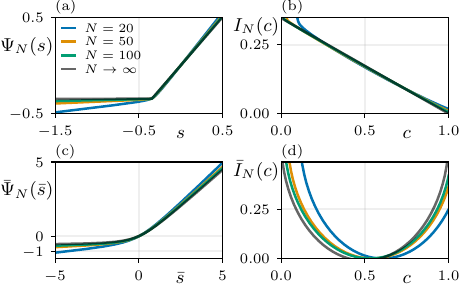}
    \caption{Bulk-Chain RW: SCGFs $\Psi_N(s)$ in (a) and rate functions $I_N(c)$ in (b) at increasing values of $N$ (coloured curves) along with their limiting functions $\Psi(s)$ and $I(c)$ (dark grey curves). Re-scaled versions $\bar{\Psi}_N(\bar{s})$ and $\bar{I}_N(c)$ (coloured curves), along with their limits $\bar{\Psi}(\bar{s})$ and $\bar{I}(c)$ (dark grey curves) in (c) and (d).}
    \label{fig4:scgf}
\end{figure}

As $N$ increases, the RW spends more time in the bulk of the graph. The ergodic value, denoted by $c^*$, can be calculated and is given by $c^* = (-18 + 23N - 8N^2 + N^3)/(N(10 - 5N + N^2))$. In large graphs, the RW tends to get lost in the bulk for a simple entropic reason: the higher the number of neighbors, the harder to find the dangling chain. Fluctuations of this model were already studied and we refer to~\cite{Carugno2022} for details on the calculations. In the panels (a) and (b) of Fig.\ \ref{fig4:scgf} we report the SCGF $\Psi_N(s)$, the rate function $I_N(c)$, and their limiting behaviour for $N \rightarrow \infty$. Evidently, a kink appears at $s_c \coloneqq -(\ln 2)/2$. Similarly to the previous case, for $s<s_c$, the RW favours the dangling chain, while for $s>s_c$, it favours the bulk. At the critical value, it splits its time equally between both phases.

For this particular model there are two distinct ways the RW can split its time between bulk and dangling chain. On the one hand, the RW could keep hopping back and forth from the node of degree $2$ and the gateway. On the other hand, it could spend some time in the dangling chain, then hop to the bulk and spend some time there before jumping back. Although both mechanisms are possible, the latter is more probable than the former. We checked this both estimating the probability of the two different events per unit time (as suggested in \cite{Whitelam2021}) and by running simulations of the driven process at $s_c$ for increasingly larger values of $N$. 

We find that trajectories are intermittent in this case too. As they are qualitatively equivalent to the previous case we refer back to Fig.\ \ref{fig2:2int} for illustration purposes. The larger the $N$ the longer the waiting time of the RW before it visits the other phase. By analytically expanding \eqref{eq:ResPsi} in $N$ as mentioned for the previous example we calculate $\tau(N)=N$ and the re-scaled function $\bar{\Psi}(\bar{s})$ (its form is lengthy and not reported here, see~\cite{Carugno2022}). We plot the latter, its Legendre transform, and numerical realisations for finite $N$ given by \eqref{eq:ResPsi} and \eqref{eq:ResI} in the panels (c) and (d) of Fig.\ \ref{fig4:scgf}. Numerical realisations smoothly approach $\bar{\Psi}(\bar{s})$ endorsing the idea that at the relevant time scale $\tau(N)$ no critical behaviour emerges.

Eventually, in Fig.\ \ref{fig3:mwt}(b), we compare the form $\tau(N) = N$---corresponding to the relaxation time scale of the driven process in this case too---with numerical simulations of the driven process waiting time at the critical value $s_c$ for different values of $N$. Numerical simulations strongly support the theory, providing further evidence that the waiting time is the key factor leading to a diverging time scale and dynamical phase coexistence.

\textit{Example 3: Three-state Random Walk.} We consider a novel model of a RW over three states named `chain1', `chain2', and `bulk', characterised by the following transition matrix:
\begin{equation}
    \label{eq:3StateTrans}
    \Pi = 
    \begin{pmatrix}
    0 & 1 & 0 \\
    \frac{1}{2} & 0 & \frac{1}{2} \\
    0 & \frac{\bar{k}}{\ln N} & 1-\frac{\bar{k}}{\ln N}
    \end{pmatrix} \, ,
\end{equation}
with $\bar{k}$ and $N$ being two positive parameters of the model such that $\bar{k} \leq \ln N$. Like the previous models, we can visualise the underlying graph as a dangling chain (composed by `chain1' and `chain2' states) linked to a bulk, which in this case is composed of a single self-looped state. Just like the Bulk-Chain RW model, increasing the value of $N$ reduces the probability for the random walker to transition from the bulk to the dangling chain. However, due to the logarithmic dependence on $N$ (which will be motivated in the following), this transition is relatively easier compared to the Bulk-Chain RW model. Finally, the RW collects also in this case an observable of the form \eqref{eq:Obs} with $f(\text{chain1}) = 1$, $f(\text{chain2}) = 2$, and $f(\text{bulk}) = \bar{k}$.

\begin{figure}
    \centering
    \includegraphics[width=\linewidth]{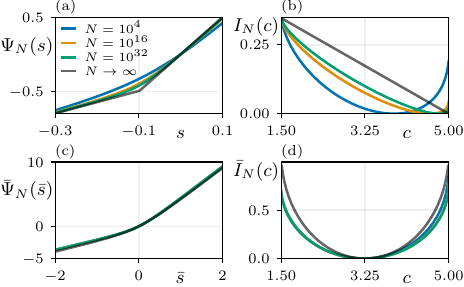}
    \caption{Three-state RW: SCGFs $\Psi_N(s)$ in (a) and rate functions $I_N(c)$ in (b) at increasing values of $N$ (coloured curves) along with their limiting functions $\Psi(s)$ and $I(c)$ (dark grey curves). Re-scaled versions $\bar{\Psi}_N(\bar{s})$ and $\bar{I}_N(c)$ (coloured curves), along with their limits $\bar{\Psi}(\bar{s})$ and $\bar{I}(c)$ (dark grey curves) in (c) and (d).}
    \label{fig6:scgf}
\end{figure}

The long-time behaviour of the observable $C_n$ is given by $c^* = (\bar{k}(5 + \ln N))/(3 \bar{k} + \ln N)$ and fluctuations can be studied with large deviation theory as outlined above. The SCGF $\Psi_N(s)$ in \eqref{eq:SCGF} can be calculated analytically but its form is lengthy and since it is not useful here is not displayed. Its limit for $N \rightarrow \infty$ in \eqref{eq:LimitN} can also be calculated and reads
\begin{equation}
    \label{eq:3StateLimiting}
    \Psi(s) = 
    \begin{cases*}
      \frac{3s - \ln 2}{2} & if \, $s < s_c$ \\
      \bar{k} s & if \, $ s > s_c$
    \end{cases*} \, ,
\end{equation}
with a kink at 
\begin{equation}
    \label{eq:Critical3State}
    s_c = \frac{\ln 2}{3 - \bar{k}} \, .
\end{equation} 
These functions, along with their transforms, i.e., $I_N(c)$ and $I(c)$, are displayed in the panels (a) and (b) of Fig.\ \ref{fig6:scgf}. Once again, when $s>s_c$, the RW will spend more time in the bulk ($c > c^*$), while for $s<s_c$ the RW favours the dangling chain ($c < c^*$).

\begin{figure}[h!]
    \centering
\includegraphics[width=\linewidth]{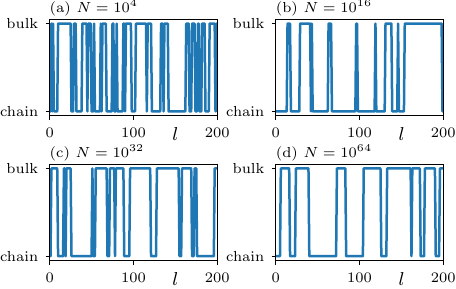}
    \caption{From (a) to (d), intermittent trajectories visiting `bulk' and `chain' at increasing values of $N$ for the three-state RW model. The larger the $N$ the longer the waiting time to visit the other phase, although because of the logarithmic scale the change is less evident with respect to the two-state RW of Fig.\ \ref{fig2:2int}.}
    \label{fig7:traj}
\end{figure}

At the critical value $s_c$ the RW will spend half its time in the dangling chain and the other half in the bulk. Even in this case intermittent behaviour arises as leading mechanism to generate fluctuations around the critical point $s_c$. We check this numerically by simulating the driven process \eqref{eq:DrivenProcess} for increasingly larger values of $N$ and plot the trajectories in Fig.\ \ref{fig7:traj}.
Even in this scenario, as the value of $N$ increases, the RW experience longer waiting times before transitioning to the other phase. 

We derive $\bar{\Psi}(\bar{s})$ as leading-in-$N$ behaviour of \eqref{eq:ResPsi} and the time scale $\tau(N)$. Analytically, we proceed by considering the characteristic equation for the dominant eigenvalue $\lambda = e^{\Psi_N(s)}$ of the tilted matrix \eqref{eq:TiltedMatrix}, i.e.,
\begin{equation}
    \label{eq:Characteristic3State}
    (e^{\bar{k}s} - \lambda)(e^{3s}-2 \lambda^2) - \frac{e^{\bar{k} s} \ln \bar{k} }{\ln N} \left( e^{3s} + e^{2 s} \lambda - 2 \lambda^2 \right) = 0 \, .
\end{equation}
We replace $s$ and $\Psi_N(s)$ as functions of $\bar{s}$ and $\bar{\Psi}_N(\bar{s})$ inverting the relations given in \eqref{eq:Ress} and \eqref{eq:ResPsi} respectively. Then, we make an educated guess and also replace $\tau(N) = \left( \ln N \right)^{\alpha}$ with $\alpha > 0$ and keep only the first order in $(\ln N)^{-\alpha}$ of \eqref{eq:Characteristic3State}. The value of $\alpha=1/2$---the smallest possible---and the smooth function $\bar{\Psi}(\bar{s})$ are found imposing that the zero of the leading term of \eqref{eq:Characteristic3State} is not the kinked function in \eqref{eq:3StateLimiting}. Figures (c) and (d) of Fig. \ref{fig6:scgf} display the plots of $\bar{\Psi}(\bar{s})$ and $\bar{I}$, along with their finite-$N$ approximations \eqref{eq:ResPsi} and \eqref{eq:ResI} respectively. The smoothness of the limiting functions suggests, once again, that there is a proper time scale in the large deviations such that no critical behaviour is observed.

Eventually, we compare in Fig.\ \ref{fig3:mwt}(c) the scaling $\tau(N) = \left( \ln N \right)^{1/2}$ with the mean waiting time computed numerically by simulating driven processes with fixed $s=s_c$ for increasingly larger values of $N$. 
Once again, $\tau(N)$ corresponds to the relaxation time scale of the driven process and its form well matches numerical simulations of the waiting time multplied by an $N$-independent prefactor. We conclude that the waiting time is the leading diverging time scale supporting phase coexistence in the first-order DPT.

\textit{Implications for Erd\"{o}s--R\'{e}nyi random graphs.} We have discussed three models that aim to capture the characteristics of a simplified version of an unbiased RW exploring an Erd\"{o}s--R\'{e}nyi (ER) random graph. An ER graph is created by randomly connecting a fixed number $N$ of nodes. The probability of connecting two nodes is determined by $\bar{k}/N$, where $\bar{k}$ represents the average connection of the graph. After generating the graph, only its largest connected component is retained and used as a base structure for a RW collecting a cost as in \eqref{eq:Obs}.

Large deviation theory has recently been employed to study this model~\cite{DeBacco2016,Coghi2019}. In the case where $\bar{k}$ is sufficiently small, indications of a DPT have been identified between a phase characterised by bulk delocalisation and another phase where the RW localises along dangling chains. However, no formal proof has been provided thus far for the existence of this DPT in the infinite-size ER graph ensemble.

The two-state and the Bulk-Chain RW models had previously been introduced~\cite{Carugno2022} as analytical models aimed at explaining the delocalisation-localisation DPT in ER graphs. Our three-state RW takes a further step in this direction. In the Bulk-Chain RW, increasing the size of $N$ expands the bulk by adding fully connected nodes. However, this does not accurately reflect the behaviour of adding a node in an ER graph, where the new node is not fully connected but, on average, maintains the same number of connections $\bar{k}$. Nevertheless, the presence of the new node increases the distance between a random node in the bulk and a node with degree $1$ in a dangling chain. Since the average shortest distance between any two nodes of a random graph increases logarithmically with $N$, our third model considers the probability of transitioning from the bulk to the dangling chain to be inversely proportional to $\ln N$. Meanwhile, we maintain a constant cost \eqref{eq:Obs} accumulated by the RW while visiting the bulk phase, which is equal to $\bar{k}$.

Indeed, this new model aligns more closely with the exploration of an ER random graph. Our three-state RW exhibits a delocalisation-localisation DPT that occurs at the critical value \eqref{eq:Critical3State}, which is inversely proportional to $\bar{k}$. This is expected in the ER graph too: the larger the $\bar{k}$ the more connected the bulk, the smaller the tilting $s$ to escape it. However, what is particularly striking about this transition is its speed, which goes like $(\ln N)^{1/2}$ and indicates the RW undergoes an exceptionally slow change in behaviour as $N$ increases.

\textit{Conclusion.} We have investigated the appearance of first-order dynamical phase transitions in large deviation functions of simple discrete-time and space dynamical systems. The key finding of our study is that a kink in the scaled cumulant generating function of a time-additive observable for a random walk indicates phase coexistence when the dominant diverging time scale is the waiting time for the random walk to transition between phases. The characterisation of such a time scale allows us to properly re-scale large deviation functions and therefore rule out non-ergodic behaviour. We have shown this with three illustrative examples that work towards a better understanding of a potential delocalisation-localisation dynamical phase transition in random walks on Erd\"{o}s--R\'{e}nyi random graphs. Our results suggest that such a transition may appear at a remarkably slow rate, scaling as $O((\ln N)^{1/2})$, which poses significant challenges for numerical studies. Finally, we note that our work could serve as a theoretical ground to accelerate the learning process in numerical sampling schemes of large deviations near dynamical phase transitions~\cite{Nemoto2017,Ferre2018,Coghi2022,Yan2022}.

\textit{Acknowledgements.} The authors are grateful to Supriya Krishnamurthy for support during the work. FC thanks Stephen Whitelam for insightful discussions at the Lawrence Berkeley National Lab and Giorgio Carugno and Hugo Touchette for valuable feedback on a first draft of the paper.

\bibliography{mybib}

\end{document}